# Exceptionally high Verdet constant in gold nanodisc arrays

*Gajendra Mulay*[1,2], *Shraddha Choudhary*[1,3], *Ashwin A Tulapurkar*[4], *Daria O Ignatyeva*[5], *Vladimir I Belotelov*[5], *Shriganesh S Prabhu*[1], and *Venu Gopal Achanta*[1,6]

[1]Department of Condensed Matter Physics and Materials Science, Tata Institute of Fundamental Research, Mumbai 400005, India.

[2]Center for Research in Nanotechnology and Science, Indian Institute of Technology Bombay, Mumbai 400076, India.

[3]Institute of Physics and Center for Nanotechnology, University of Münster, Münster 48149, Germany.

[4]Department of Electrical Engineering, Indian Institute of Technology Bombay, Mumbai, 400076, India.

[5]Faculty of Physics, Lomonosov Moscow State University, Moscow, 119992, Russia.

[6]On lien at CSIR-National Physical Laboratory, Dr K.S. Krishnan Marg, New Delhi 110012, India.

Email Address: [1,2]gajendra@tifr.res.in, [1,6]achanta@tifr.res.in.

Magneto-optical effects in non-magnetic noble metals can be greatly enhanced by leveraging the in-plane Lorentz force at engineered plasmonic resonances. We demonstrate a 2-D array of gold nanodiscs designed to host a hybrid resonance of localized plasmon and surface lattice modes. The structure exhibits a Verdet constant of $1.98 \times 10^6$ °·T$^{-1}$m$^{-1}$, corresponding to a Faraday rotation of -0.15° at a 1 T magnetic field. This Verdet constant represents a 15-fold enhancement over unpatterned gold and is highly competitive with many plasmon-enhanced diamagnetic nanostructures. These findings offer new opportunities for harnessing strong magneto-plasmonic effects in optoelectronic devices by patterning common non-magnetic metals.

## Introduction

Magneto-optical (MO) effects arising from the breaking of time-reversal symmetry in materials are manifested through the off-diagonal elements of the dielectric constant tensor[1]. The strength of the Faraday effect is quantified by the Verdet constant, which is exceptionally high in specialized systems like alkali gases or materials such as Terbium Gallium Garnet (TGG)[2]. In naturally occurring materials, the Verdet constant typically ranges from $10^3$–$10^4$ °·$T^{-1}m^{-1}$ for inorganics to $10^4$–$10^6$ °·$T^{-1}m^{-1}$ for organic polymers[3]. While the enhancement of MO effects in transparent ferromagnetic materials using plasmons, Mie resonances, or metasurfaces[4-15] is well-established, the weaker response in non-magnetic noble metals is also a subject of significant investigation[16-18]. Inducing a strong MO effect in materials like gold, which are standard for electrical contacts in optoelectronics, remains a key opportunity for designing novel active devices. The key challenge is in enhancing the weak non-diagonal polarizability ($\alpha_{xy}$) of the metal that is related to the off-diagonal elements of the dielectric tensor. Non-diagonal polarizability is proportional to the applied magnetic field, which limits the possibilities to enhance MO effects, as the magnetic fields needed to see large MO effects is prohibitively high, exceeding 10s of Tesla. An alternative is to enhance the $\alpha_{xy}$ at a designed resonance in a structured metal nanostructure.

Hybrid resonances, formed by coupling Localized Surface Plasmon Resonances (LSPR) with Surface Lattice Resonances (SLR), are known to produce sharp, high-quality factor optical responses and have been studied to demonstrate novel functionalities like enhancement of photocurrent[19,20]. In this work, we propose and demonstrate a gold nanodisc array that supports a hybridized LSPR-SLR mode to significantly enhance the Faraday effect. The proposed structure enhances the effect in two ways. The coupling of individual LSPR modes into a collective SLR creates a strong, narrow resonance in the Faraday spectrum. Additionally, the nanostructuring itself modifies the gyrotropic part of the permittivity of gold nanodiscs, which also amplifies the magneto-optical response. At this resonance, we measured a Verdet constant three orders of magnitude higher than in other non-magnetic metal nanostructures. This gives a total Faraday rotation that is 15 times greater than an unpatterned gold film of the same thickness.

In the following, we describe the nanostructure supporting hybridized LSPR and SLR resonances and present the details of the 2-D array of gold nanodiscs. We carried out detailed numerical studies and optical transmission measurements to understand the dispersion of three structures with different lattice constants each. The magneto-optical measurements, as well as numerical simulation results, are then presented to show the magneto-optical Faraday rotation, Verdet constant, and the Figure of Merit. We compare these with literature values for diamagnetic materials and nanostructures. Both the Verdet constant and Faraday rotation angle show giant enhancement compared to those reported in nanostructures with non-magnetic metals, demonstrating the potential of resonant structures designed for specific applications.

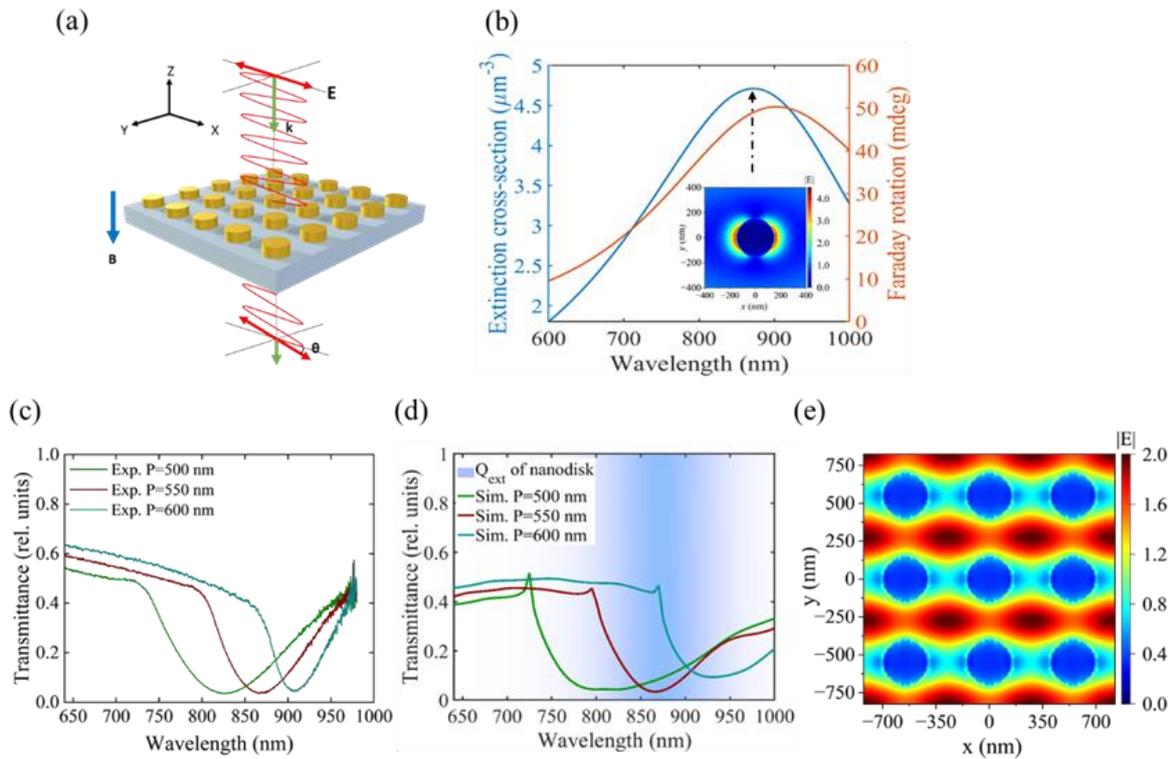

*Fig. 1.* **Surface Lattice Mode** *(a) Schematic of an array of gold nanodiscs and the geometry of the applied optical and external magnetic fields. (b) The extinction cross-section and Faraday rotation spectrum of a single Au nanodisc were calculated using modified long-wavelength*

*approximation (MLWA)[21]. The inset shows the electric field distribution at 860 nm resonant wavelength calculated at the centre of the disc using FDTD. (c) Experimental and (d) simulated transmittance spectra of the nanodisc arrays. The blue-shaded region shows the extinction cross-section ($Q_{ext}$) for comparison. (e) Simulated electric field profile for 550 nm period gold nanodisc array showing the lattice mode excited by light polarized along the x-axis.*

**Structure supporting hybridized LSPR and SLR:** We study a nanostructure consisting of gold nanodiscs with a diameter $D = 300$ nm and height $h = 80$ nm arranged in a 2-D array with period, $P$=500 nm, 550 nm, and 600 nm on a quartz substrate (Fig. 1a). Gold nanodisc arrays were fabricated by a multi-step process involving sputtering of gold thin films, electron beam lithography for nanopatterning, and reactive ion etching for transferring the pattern to the metal film. The periods were chosen to excite the coupled LSPR and SLR modes.

In metal nanodiscs, excitation of LSPR significantly modifies the effective polarizability tensor[22], so the extinction cross-section $Q_{ext} = \frac{4}{\pi D^2} k \, \text{Im}[\alpha_{xx}]$ has a maximum at a wavelength depending on the nanodisc diameter $D$ and height $h$ (blue curve in Fig.1b). This maximum is associated with excitation of an electric dipole in the nanodisc (inset of Fig.1b). Moreover, in the vicinity of this resonance condition, an electrical dipole, transversal to that excited by the induced light is also generated under the impact of the external magnetic field directed along the axis of the nanodisc, which leads to the appearance of the off-diagonal $\alpha_{xy}$ polarizability component. This results in the substantial broadband increase of the Faraday rotation $\theta_F = \text{Re}\left(\frac{\alpha_{xy}}{\alpha_{xx}}\right)$ in the vicinity of LSPR (orange line in Fig.1b).

While LSPR in nanoparticles enhances the local field over sub-wavelength dimensions, their arrangement in 2-D plasmonic crystal lattice results in the interaction between the neighboring nanodiscs and modification of the optical and magneto-optical response. A plasmonic crystal, a periodic 2-D array, enables the coupling of diffraction modes with the plasmon resonances[20,23]. This coupling becomes essential if some diffracted wave propagates along the surface of the array[23-25], i.e. under the conditions close to the Rayleigh anomalies (RA) determined by:

$$\left(\frac{2\pi}{\lambda}\sin\theta + \frac{2\pi}{P}a_x\right)^2 + \left(\frac{2\pi}{P}b_y\right)^2 = \left(\frac{2\pi}{\lambda}n_\beta\right)^2 \qquad (1)$$

where $\lambda$ is the wavelength of the incident light, $\theta$ is the angle of incidence, $P$ is the period of the structure, $n_\beta$ represents the refractive index of the environment (either $n_{air}$ or $n_{quartz}$), and $a_x$ and $b_y$ signify distinct diffraction orders of the grating. It follows from Eq. (1) that for some angle of incidence and lattice period, Rayleigh anomaly (RA) may coincide with LSPR, which leads to the diffractive coupling of individual LSPR of nanodiscs and formation of the hybrid SLR. For normal incidence, SLR appears for $P$=550 nm, as is seen by the experimental and calculated transmittance - modification of the LSPR shape and narrowing linewidth (Fig. 1 c,d, respectively) and near field profile of the optical electric field (Fig. 1e). Figure 1e was obtained by the FDTD simulations with periodic boundary conditions in the x- and y-directions and PML in the z-direction. The lattice mode shows field maxima in the air region at the resonance wavelength of 860 nm.

To study in more detail the interaction between the LSPR in individual nanodiscs and the SLR appearing in the lattice, finite difference time domain simulations and angle-resolved white light transmission measurements are carried out. Figures 2(a-c) show the SEM images of the fabricated structures with three different periods as well as the dispersion plots that are simulated Figs. 2(d-f) and measured Figs. 2(g-i). The dashed lines in Figs. 2 (d-i) are the analytically calculated RAs for each array using Eq.(1). While keeping the diameter of the nanodisc constant, when the period of the array is varied, the spectral position of RA changes and, therefore, SLR for different periods appears at different wavelengths. For example, at normal incidence, the SLR is at 730 nm, 803 nm, and 876 nm for $P$ = 500 nm, 550 nm, and 600 nm, respectively (Figs. 2(d-i)).

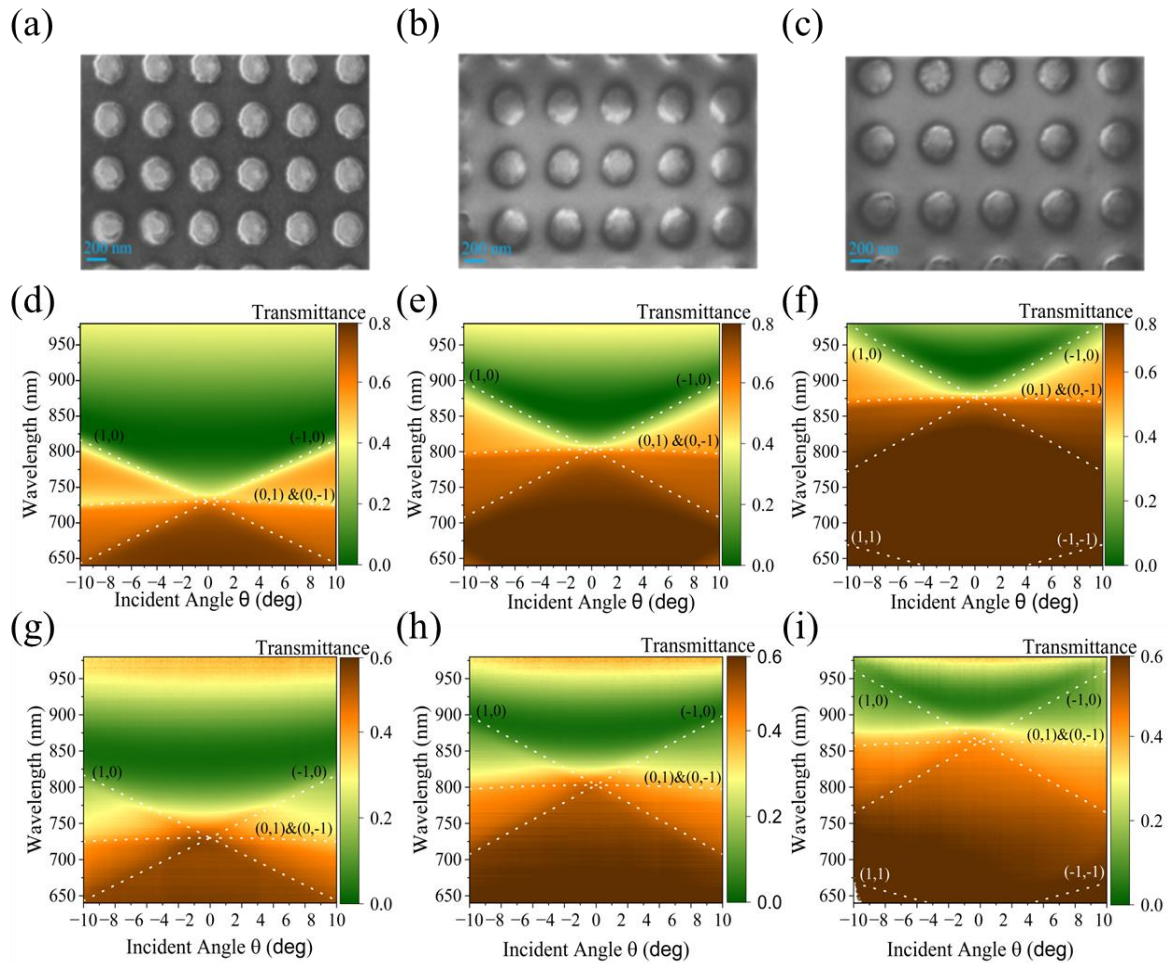

*Fig.2: **Simulated and Measured Dispersion** SEM images of different periodic structures are shown in (a-c). Dispersion plots for 3 different periods of 500, 550, and 600 nm. Simulated (d-f) and measured (g-i) transmission spectra for different incident angles. Dashed lines are the analytically calculated modes for each period.*

**Faraday rotation and Verdet constant of gold nanodisc arrays:** The resonant interaction of light with gold nanodiscs is accompanied by resonances in the Faraday rotation (Figs. 3(a-c)). Figure S1 in the Supplement presents the as-measured total optical rotation and ellipticity, which show very high values of 3.8° and -1.5°, respectively. This large field-independent optical rotation is attributed to unavoidable, minor structural anisotropies or asymmetries introduced during the multi-step nanofabrication process. While the arrays are designed to be perfectly symmetric, small deviations in the shape or placement of the nanodiscs can break this symmetry, leading to a polarization conversion[26]. This effect is strongly amplified by the high quality-factor of the surface lattice resonance, resulting in a large background rotation. This observation is supported by numerical simulations, the results of which are presented in the Supplementary Information (Fig. S5). After subtracting the field independent component and the quartz substrate contribution, the magnetic field-dependent Faraday rotation ($\theta_F$) is given by,

$$\theta_F = \theta_{TR}(B) - \theta_{TR}(0) - \theta_{F\,SiO_2}(B)$$
(2)

In this, $\theta_{TR}(B)$ is the total optical rotation through the nanodisc array at an applied magnetic field ($B$) containing both field-dependent and field-independent contributions, and the same value measured at zero magnetic field and thus corresponding to non-magnetic rotation is given by $\theta_{TR}(0)$. $\theta_{F\,SiO_2}$ is the Faraday rotation of the $SiO_2$ substrate (Figs. S1 and S2 in Supplement) at that applied field (measurement setup details are in Methods). Similarly, the measured magnetic field-dependent ellipticity angles for the 500 nm, 550 nm, and 600 nm period arrays are shown in Fig. S2 (Supplement).

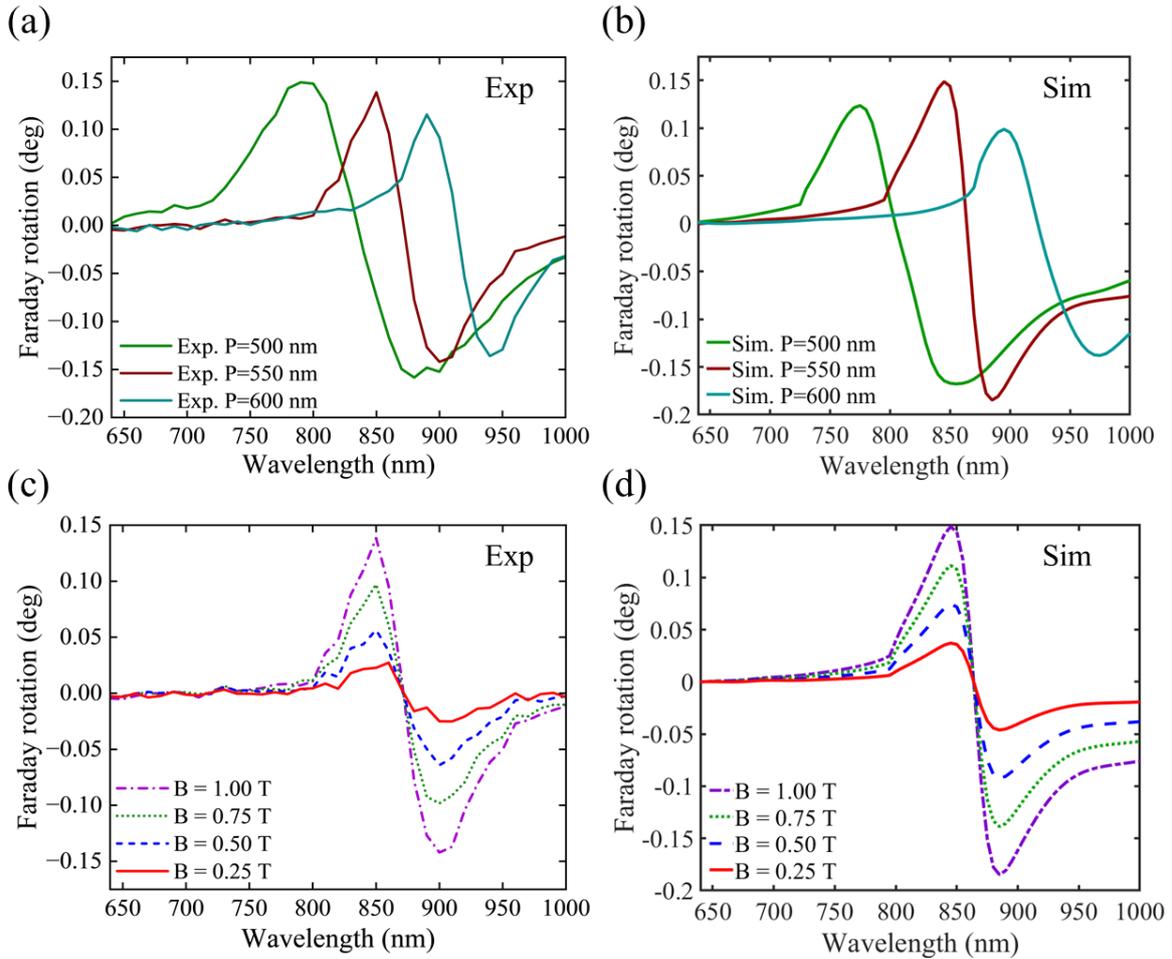

*Fig. 3: **Faraday Rotation** (a) Measured and (b) simulated Faraday rotation for the three nanodisc arrays with different periods of 500 nm, 550 nm, and 600 nm, respectively, for the external magnetic field B=1T. The experimental Faraday rotation values plotted here are after subtracting the field-independent and quartz substrate contributions; (c) Measured and (d) simulated Faraday rotation for different external magnetic field values (see the legends) for P=550 nm nanodisc array.*

The Faraday rotation of the Au nanodisc arrays exhibits a prominent resonant enhancement near the SLR resonance. Specifically, the excitation of SLR mode in the Au nanodisc resonators results in a Fano line shape in the Faraday rotation spectra. For instance, a Faraday angle peak of 0.15° is observed at λ = 790 nm, and a dip of −0.15° is observed at λ = 880 nm for a lattice period P=500 nm at B=1T. Such a large Faraday rotation is caused by the interplay of two different factors. First, SLR in nanodisc arrays produces resonances narrower than LSPR of individual discs (compare Fig. 1b and Figs. 1c,1d)[27]. The interplay of the LSPR and SLR resonances results in the Fano-type shape of the Faraday rotation resonance and allows the enhancement of the observed Faraday rotation. However, here we face the second important mechanism of the Faraday effect enhancement related to the nanostructuring of Gold. In diamagnetic metals, off-diagonal elements of the permittivity tensor responsible for the Faraday rotation are determined by the Lorentz force acting on free electrons. In the case of nanostructures, the surface charge carriers play an important role in their magnetism, whose properties obviously differ from those of bulk electrons. Recent studies showed a giant enhancement in the magnetic susceptibility of nanostructured gold due to the surface effects[28,29]. As the magnetic susceptibility increases, the internal magnetic field in gold nanodiscs is modified, and, as a result, gyrotropic terms of the permittivity tensor are enhanced[28,29]. We observed a tenfold enhancement of the real part of the off-diagonal permittivity tensor, which for the nanostructure was $\varepsilon_{xy} = 0.69 - 0.20i$ at 850 nm wavelength (see Methods for the details). Both reasons lead to a significant enhancement of the Faraday effect in the nanostructured gold, as we observed here. For the smooth Au film of the same thickness ($h$ =80 nm), the Faraday rotation measured is 0.01° only. Thus, the 2-D nanodisc array provides a 15-times enhancement of the Faraday polarization rotation that arises due to the resonance effect in the structure (Fig. S1). If one considers the reduction in the amount of gold during patterning, equal to $\pi D^2/4P^2 \approx 0.25$, the enhancement of the Faraday rotation becomes even more striking.

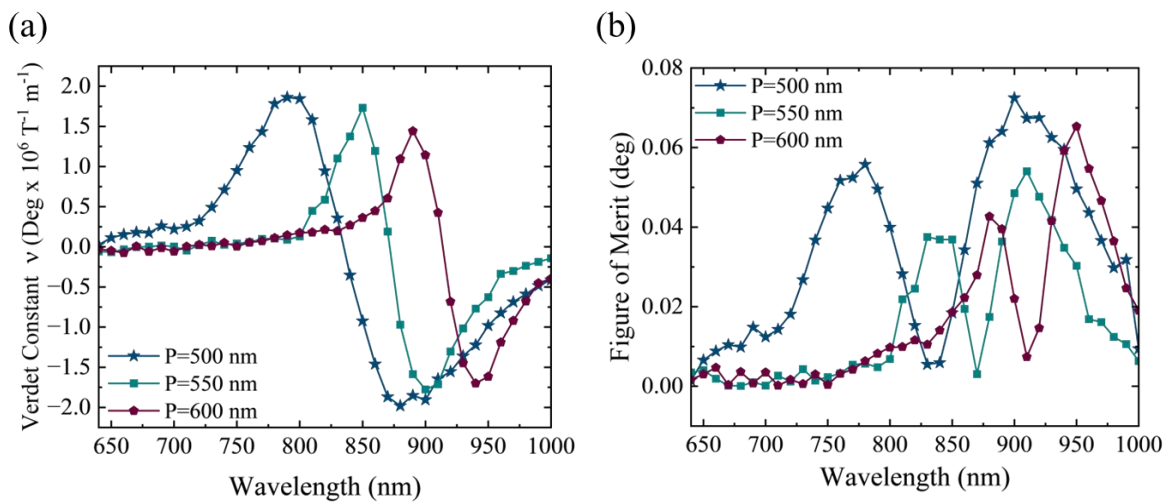

*Fig. 4. **Verdet Constant and FOM** For different lattice periods (a) Verdet constant as a function of wavelength and (b) Figure of Merit (FOM). Values were calculated using the magneto-optical Faraday rotation.*

As expected, the Faraday rotation demonstrates a linear dependence on the applied magnetic field, exhibiting a pronounced resonant behavior in proximity to the SLR wavelength, as illustrated in Figure 3c and further detailed in Figure S4 of the Supplement. In order to quantify this magneto-optical response more explicitly, we performed a linear fit of the Faraday rotation as a function of the magnetic field for various wavelengths. From the slope of these fits, we successfully extracted the Verdet constant ($v$) from,

$$\theta_F = vBl \qquad (3)$$

where $l$ is the effective distance (height of the Au nanodisc resonators, 80 nm). For the three different lattice periods, the Verdet constant at different wavelengths is presented in Fig. 4(a). Among the various structures, the one with a lattice period of 500 nm exhibited the highest Verdet constant of $1.98 \times 10^6 \,°.T^{-1}m^{-1}$ at 880 nm. As shown in the Table 1, this value represents a significant enhancement for a plasmonic system and is orders of magnitude higher than that of many other non-magnetic materials.

Table 1: The Verdet constant ($v$) reported in different non-magnetic materials and structures with linear MO response.

| Material description | Origin of Faraday rotation | $v(\lambda)$ ($\times 10^4 \text{degT}^{-1}\text{m}^{-1}$) | Reference |
|---|---|---|---|
| Chiral Polyfluorene | π-conjugation / B term | -38 (@410 nm) | 30 |
| Phthalocyanine | A term | -33 (@800 nm) | 31 |
| Au NPs/BK7 glass | LSPR | 28.98 (@532nm) | 32 |
| Au NPs colloidal solution | SPR | 0.24 (@ 575 nm) | 33 |
| Au NP array on quartz (P=500 nm; fill factor 28.2%) | LSPR + SLR | 186 (@790nm), -198 (@880nm) | This study |
| Au NP array on quartz (P= 550 nm; fill factor 23.3%) | LSPR + SLR | 173 (@850nm), -177 (@900nm) | This study |
| AuNP array on quartz (P= 600 nm; fill factor 19.6%) | LSPR + SLR | 144 (@890nm), -169 (@940nm) | This study |

Besides the high Verdet constant and Faraday rotation, an important point from a practical point of view is the value of the figure of merit (FOM) at specific wavelengths (Fig.4(b)). The FOM is defined as $|\theta_F|T^{0.5}$, where $\theta_F$ is the magneto-optical Faraday rotation and $T$ is the transmittance[34,35]. Fano-shaped Faraday resonance is clearly observed. The maximum magnitudes of the Faraday rotation (Figs. 3a, 3b) are spectrally shifted from the SLR resonance center where transmittance is at a minimum (Fig. 1c, d). This allows one to avoid situations where the enhancement of the Faraday rotation is accompanied by a significant decrease in transmittance, which is a typical situation for resonant plasmonic structures. Using the field-dependent Faraday rotation, the structure with $P$=500 nm yields a FOM of 0.07° with a simultaneous transmittance of 0.23. This value is close to what is usually obtained not only in non-magnetic[32], but also in magnetic plasmonic structures[6, 36-43].

## Conclusion

In this work, we have demonstrated that a carefully designed plasmonic nanostructure composed of a non-magnetic metal can generate a prominent magneto-optical (MO) response. By fabricating a gold nanodisc array that supports a hybridized LSPR-SLR mode, we achieved a significant enhancement in Faraday rotation. Our central finding is a field-dependent Verdet constant of $1.98 \times 10^6 \,°.T^{-1}m^{-1}$. Furthermore, the structure exhibits a promising figure of merit

(FOM) of 0.07° with a simultaneous high transmittance of 0.23, indicating its potential for device applications. These values are not only order of magnitude larger than that of unstructured gold but is also comparable with many plasmon-enhanced magnetic nanoparticle systems. In summary, these results open new possibilities for leveraging strong magneto-plasmonic effects in common, non-magnetic metals, enabling the design of novel active and non-reciprocal components for integrated optoelectronic devices.

## Acknowledgments
This work was supported by the Department of Science and Technology, Government of India, under the project DST/INT/RUS/RSF/P-76/2023(G). All authors at institutions affected by sanctions have contributed to this scientific work in a personal capacity. DOI and VIB acknowledge the Russian Science Foundation project N 24-42-02008, and the Russian Quantum Centre.

## Research funding
None declared.

## Author contributions
All authors have accepted responsibility for the entire content of this manuscript and consented to its submission to the journal, reviewed all the results and approved the final version of the manuscript.

## Conflict of Interest
The authors declare no conflict of interest.

## Methods
### Fabrication:
A layer of 80 nm gold was deposited on a quartz substrate using a direct current (DC) sputtering system. A negative e-beam resist AR-N 7520.11 was spin-coated on the gold film at 4000 rpm. E-beam lithography was employed for transferring the square nanodisc array pattern onto the resist. The dose parameter was optimized to ensure high-resolution features. Subsequently, the resist was developed for 90 seconds in AR300-46 followed by a DI-water rinse for 30 seconds. Reactive Ion Etching (RIE) was utilized to etch the exposed gold region, resulting in a nanodisc pattern. The Argon plasma etching of gold was carried out for 15 minutes with an RF power of 75 W, a flow rate of 50 sccm, and a chamber pressure of 5 x $10^{-3}$ mbar in a SENTEC SI500 ICP-RIE system. The residual resist was removed by $O_2$ plasma etching for 10 minutes at an RF power of 80W and a flow rate of 50 sccm.

### Experimental Setups:
Dispersion measurement is measured by measuring the optical transmission spectra as a function of the incident angle $\theta_{in}$. For this, the sample was positioned on a motorized rotation stage. White light, generated by a halogen lamp, was near collimated on the sample using a combination of apertures and lenses. An Ocean Optics (Flame) spectrometer measured the optical transmission spectra. The measured transmission spectra were normalized with the transmission spectra of the quartz substrate for each incident angle.

For measuring the Faraday rotation and Ellipticity, a supercontinuum laser (NKT) with a tunable wavelength from 500 nm to 1700 nm was used as a light source. The input polarization of the incident light was fixed using a Glan-Thompson polarizer with a $10^5$:1 extinction ratio.

Two quartz lenses were used to focus and collect light from the sample. Magnetic fields for the measurements were generated using an electromagnet. A photoelastic modulator (PEM) is placed in the light path after the sample and is aligned to match the orientation of the input polarizer. PEM is followed by an analyzer oriented at 45° with respect to the PEM. The signal is detected by a photodiode, which is then sent for further analysis to a lock-in amplifier and digital multimeter [44]. Supplementary Fig. S6 has the Faraday rotation measurement results of quartz substrate.

**Numerical Simulations:**

The FDTD method (Lumerical FDTD solutions) and Rigorous coupled-wave analysis (RCWA) method were employed to calculate the optical scattering, transmission spectra, and Faraday rotation and electromagnetic field distributions.

To obtain the transmission spectra as a function of the incident angle $\theta_{in}$, periodic boundary conditions were applied along the x- and y-directions, while PMLs (Perfectly Matched Layers) were utilized along the z-direction. A Broadband Fixed Angle Source Technique (BFAST) employing a plane wave source was utilized, where the electric field was oriented along the x-axis. The simulations utilize a fine grid, with each element measuring 10 nm. The electric fields ($|E|$) were recorded in both the x-z and the x-y planes at respective resonance wavelengths. A perfectly matched layer (PML) boundary condition along the x, y, and z directions, along with a total-field scattering field source, was used to obtain the optical properties of a single gold nanodisc.

The diagonal and off-diagonal permittivity of gold was taken into account using the Drude-Lorentz model:

$$\varepsilon_{xx} = 1 + \frac{i\omega_p^2(\gamma - i\omega)}{\omega((\gamma - i\omega)^2 + \omega_c^2)};$$

$$\varepsilon_{xy} = \frac{i\omega_p^2 \omega_c}{\omega((\gamma - i\omega)^2 + \omega_c^2)};$$

with the parameters $\omega_p = 7.8 [\text{eV}]$, $\gamma = 0.1 [\text{eV}]$, $\omega_c = 2.06 \cdot 10^{-3} B$ [eV T$^{-1}$] reported in Ref. 21 To match the experimentally measured gold properties, Johnson and Christy [45] values were tuned to get a good agreement with the experimentally measured rotation for a smooth gold film of 0.01 deg at 850 nm wavelength $\varepsilon_{xy} = -0.07 - 0.39i$.

We observed a significant enhancement of the magneto-optical properties due to the nanopatterning, such that for the nanostructures $\varepsilon_{xy}^{nano} = 10\text{Re}(\varepsilon_{xy}) + i \cdot \text{Im}(\varepsilon_{xy})$. The value for or the wavelength of 850 nm is $\varepsilon_{xy} = 0.69 - 0.20i$.

**References**

1. A. K. Zvezdin, and V. A. Kotov, *Modern Magnetooptics and Magnetooptical Materials* (CRC Press, Boca Raton, FL 1997**)**.
2. K. J. Carothers, R. A. Norwood, and J. Pyun, *Chem. Materials* **34**, 2531–2544 (2022).
3. Z. Nelson, L. Delage-Laurin, and T. M. Swager, Journal of the American Chemical Society 144 (27), 11912-11926 (2022).
4. V. I. Belotelov, I. A. Akimov, M. Pohl, et al., *Nature Nanotech.* **6**, 370–376 (2011).
5. V. I. Belotelov, L. E. Kreilkamp, I. A. Akimov, et al., *Nat. Commun.* **4**, 2128 (2013).
6. J. Y. Chin, T. Steinle, T. Wehlus, et al., *Nat. Commun.* **4**, 1599 (2013).
7. S. Xia, D. O. Ignatyeva, Q. Liu, et al., *Laser Photon. Rev.* 16, 2200067 (2022).

8. S. Xia, D. O. Ignatyeva, Q. Liu, et al., *ACS Photonics* **9**, 1240–1247 (2022).
9. V.I. Belotelov D. A. Bykov, L. L. Doskolovich, et al., *Opt. Lett.* **34**, 398–400 (2009).
10. V. I. Belotelov, L. E. Kreilkamp, A. N. Kalish, et al., *Phys. Rev. B* **89**, 045118 (2014).
11. Z. Wu, M. Levy, V. J. Fratello, et al., *Appl. Phys. Lett.* **96**, 051125 (2010).
12. M. G. Barsukova, A. S. Shorokhov, A. I. Musorin, et al., *ACS Photonics* **4**, 2390–2395 (2017).
13. A. N. Kalish, R. S. Komarov, M. A. Kozhaev, et al., *Optica* **5**, 617–623 (2018).
14. M. Pohl, L. E. Kreilkamp, V. I. Belotelov, et al., *New J Phys*. **15**, 075024 (2013).
15. D. M. Krichevsky, A. N. Kalish, M. A. Kozhaev, et al., *Phys. Rev. B* **102**, 144408 (2020).
16. G. M. Manera, A. Colombelli, A. Taurino, et al., Sci Rep 8, 12640 (2018).
17. A. Gabbani, G. Petrucci, and F. Pineider, J. Appl. Phys. 129(21): 211101 (2021).
18. J. Foxley, and K. L. Knappenberger, Annu Rev Phys Chem. 74:53-72 (2023).
19. D. Jalas, A. Petrov, M. Eich, et al., *Nature Photon.* **7**, 579–582 (2013).
20. V. G. Kravets, A. V. Kabashin, W. L, Barnes, et al., *Chem. Rev*. **118**, 5912–5951 (2018).
21. A. K. González-Alcalde, X. Shi, R. B. Wilson, et al., *J. Opt. Soc. Am. B* **41**, 2480–2487 (2024).
22. B. Sepúlveda, J. B. González-Diaz, A. Garcia-Martin, et al., *Phys. Rev. Lett.* **104**, 147401 (2010).
23. B. Auguié, X. M. Bendana, W. L. Barnes, et al., *Phys. Rev. B* **82**, 155447 (2010).
24. S. Murai, M. A. Verschuuren, G. Lozano, et al., and J. Gómez Rivas, *Opt. Express* **21**, 4250–4262 (2013).
25. B. B. Rajeeva, L. Lin, and Y. Zheng, *Nano Research* **11**, 4423–4440 (2018).
26. A. V. Baryshev, H. Uchida, and M. Inoue, J. Opt. Soc. Am. B 30, 2371-2376(2013).
27. M. Kataja, T. K. Hakala, A. Julku, et al., *Nat. Commun.* **6**, 7072 (2015).
28. S. Trudel, *Gold Bull* **44**, 3–13 (2011).
29. A. Hernando, A. Ayuela, P. Crespo, et al., *New J. Phys.* **16**, 073043 (2014).
30. C. K. Lim, M. J. Cho, A. Singh, et al., Nano Lett. 16 (9), 5451−5455 (2016).
31. Z. Nelson, L. Delage-Laurin, M. D. Peeks, et al., J. Am. Chem. Soc. 143 (18), 7096−7103(2021).
32. H. Zhu, M. Gao, C. Pang, et al., *Small Science,* **2**, 2100094 (2022).
33. O. H. C. Cheng, D. H. Son, et al., Nature Photonics 14, 365-368 (2020).
34. F. Royer, B. Varghese, E. Garnet, et al., *ACS Omega* **5**, 2886–2892 (2020).
35. A. B. Khanikaev, A. V. Baryshev, A. A. Fedyanin, et al., *Opt. Express* **15**, 6612 (2017).
36. D. Floess, M. Hentschel, T. Weiss, et al., Phys. Rev. X 7, 021048 (2017).
37. A. Lopez-Ortega, M. Takahashi, S. Maenosono, et al., Nanoscale 10, 18672-18679(2018).
38. N. Maccaferri, L. Bergamini, M. Pancaldi, et al., Nano Lett. 16, 2533-2542 (2016).
39. H. Uchida, Y. Masuda, R. Fujikawa, et al., *J. Magn. Magn. Mater.* **321**, 843–845 (2009).
40. H. Uchida, Y. Mizutani, Y. Nakai, et al., *J. Phys. D* **44**, 064014 (2011).
41. A. V. Baryshev, and A. M. Merzlikin, *J. Opt. Soc. Am. B* **33**, 1399–1405 (2016).
42. F. Zhang, T. Atsumi, X. Xu, et al., *Nanophotonics* **11**, 275–288 (2021).
43. K. Sato, and T. Ishibashi, *Front. Phys.* **10**, 946515 (2022).
44. S. Vandendriessche, and T. Verbiest, *Polarization Science and Remote Sensing VI* 88730Z SPIE (2013).
45. P. B. Johnson, and R. W. Christy, *Phys. Rev. B* **6**, 4370–4379 (1972).

# References with Titles

1. A. K. Zvezdin, and V. A. Kotov, *Modern Magnetooptics and Magnetooptical Materials* (CRC Press, Boca Raton, FL 1997**)**.
2. K. J. Carothers, R. A. Norwood, and J. Pyun, "High Verdet Constant Materials for Magneto-Optical Faraday Rotation: A Review," *Chem. Materials* **34**, 2531–2544 (2022). https://doi.org/10.1021/acs.chemmater.2c00158
3. Z. Nelson, L. Delage-Laurin, and T. M. Swager, "ABCs of Faraday Rotation in Organic Materials," Journal of the American Chemical Society 144 (27), 11912-11926 (2022). https://doi.org/10.1021/jacs.2c01983
4. V. I. Belotelov, I. A. Akimov, M. Pohl, V. A. Kotov, S. Kasture, A. S. Vengurlekar, V. G. Achanta, D. R. Yakovlev, A. K. Zvezdin, and M. Bayer, "Enhanced magneto-optical effects in magnetoplasmonic crystals," *Nature Nanotech.* **6**, 370–376 (2011). https://doi.org/10.1038/nnano.2011.54
5. V. I. Belotelov, L. E. Kreilkamp, I. A. Akimov, A. N. Kalish, D. A. Bykov, S. Kasture, V. J. Yallapragada, V. G. Achanta, A. M. Grishin, S. I. Khartsev, M. Nur-E-Alam, M. Vasiliev, L. L. Doskolovich, D. R. Yakovlev, K. Alameh, A. K. Zvezdin, and M. Bayer, "Plasmon-mediated magneto-optical transparency," *Nat. Commun.* **4**, 2128 (2013). https://doi.org/10.1038/ncomms3128
6. J. Y. Chin, T. Steinle, T. Wehlus, D. Dregely, T. Weiss, V. I. Belotelov, B. Stritzker, and H. Giessen, "Nonreciprocal plasmonics enables giant enhancement of thin-film Faraday rotation," *Nat. Commun.* **4**, 1599 (2013). https://doi.org/10.1038/ncomms2609
7. S. Xia, D. O. Ignatyeva, Q. Liu, H. Wang, W. Yang, J. Qin, Y. Chen, H. Duan, Y. Luo, O. Novák, M. Veis, L. Deng, V. I. Belotelov, and L. Bi, "Circular Displacement Current Induced Anomalous Magneto-Optical Effects in High Index Mie Resonators," *Laser Photon. Rev.* 16, 2200067 (2022). https://doi.org/10.1002/lpor.202200067
8. S. Xia, D. O. Ignatyeva, Q. Liu, J. Qin, T. Kang, W. Yang, Y. Chen, H. Duan, L. Deng, D. Long, M. Veis, V. I. Belotelov, and L. Bi, "Enhancement of the Faraday Effect and Magneto-optical Figure of Merit in All-Dielectric Metasurfaces, " *ACS Photonics* **9**, 1240–1247 (2022). https://doi.org/10.1021/acsphotonics.1c01692
9. V.I. Belotelov D. A. Bykov, L. L. Doskolovich, A. N. Kalish, V. A. Kotov, and A. K. Zvezdin, "Giant magneto-optical orientational effect in plasmonic heterostructures," *Opt. Lett.* **34**, 398–400 (2009). https://doi.org/10.1364/OL.34.000398
10. V. I. Belotelov, L. E. Kreilkamp, A. N. Kalish, I. A. Akimov, D. A. Bykov, S. Kasture, V. J. Yallapragada, V. G. Achanta, A. M. Grishin, S. I. Khartsev, M. Nur-E-Alam, M. Vasiliev, L. L. Doskolovich, D. R. Yakovlev, K. Alameh, A. K. Zvezdin, and M. Bayer, "Magnetophotonic intensity effects in hybrid metal-dielectric structures," *Phys. Rev. B* **89**, 045118 (2014). https://doi.org/10.1103/PhysRevB.89.045118
11. Z. Wu, M. Levy, V. J. Fratello, and A. M. Merzlikin, "Gyrotropic photonic crystal waveguide switches," *Appl. Phys. Lett.* **96**, 051125 (2010). https://doi.org/10.1063/1.3309715
12. M. G. Barsukova, A. S. Shorokhov, A. I. Musorin, D. N. Neshev, Y. S. Kivshar, and A. A. Fedyanin, "Magneto-Optical Response Enhanced by Mie Resonances in Nanoantennas," *ACS Photonics* **4**, 2390–2395 (2017). https://doi.org/10.1021/acsphotonics.7b00783
13. A. N. Kalish, R. S. Komarov, M. A. Kozhaev, V. G. Achanta, S. A. Dagesyan, A. N. Shaposhnikov, A. R. Prokopov, V. N. Berzhansky, A. K. Zvezdin, and V. I. Belotelov, "Magnetoplasmonic quasicrystals: an approach for multiband magneto-optical response," *Optica* **5**, 617–623 (2018). https://doi.org/10.1364/OPTICA.5.000617
14. M. Pohl, L. E. Kreilkamp, V. I. Belotelov, I. A. Akimov, A. N. Kalish, N. E. Khokhlov, V. J. Yallapragada, V. G. Achanta, M. Nur-E-Alam, M. Vasiliev, D. R. Yakovlev, K. Alameh, A.

K. Zvezdin, and M. Bayer, "Tuning of the transverse magneto-optical Kerr effect in magneto-plasmonic crystals," *New J Phys*. **15**, 075024 (2013). https://doi.org/10.1088/1367-2630/15/7/075024
15. D. M. Krichevsky, A. N. Kalish, M. A. Kozhaev, D. A. Sylgacheva, A. N. Kuzmichev, S. A. Dagesyan, V. G. Achanta, E. Popova, N. Keller, and V. I. Belotelov, "Enhanced magneto-optical Faraday effect in two-dimensional magnetoplasmonic structures caused by orthogonal plasmonic oscillations," *Phys. Rev. B* **102**, 144408 (2020). https://doi.org/10.1103/PhysRevB.102.144408
16. G. M. Manera, A. Colombelli, A. Taurino, A. G Martin, and R. Rella, "Magneto-Optical properties of noble-metal nanostructures: functional nanomaterials for bio sensing," Sci Rep 8, 12640 (2018). https://doi.org/10.1038/s41598-018-30862-3
17. A. Gabbani, G. Petrucci, and F. Pineider, "Magneto-optical methods for magnetoplasmonics in noble metal nanostructures," J. Appl. Phys. 129(21): 211101 (2021). https://doi.org/10.1063/5.0050034
18. J. Foxley, and K. L. Knappenberger, "Magneto-Optical Properties of Noble Metal Nanostructures," Annu Rev Phys Chem. 74:53-72 (2023). https://doi.org/10.1146/annurev-physchem-062322-043108
19. D. Jalas, A. Petrov, M. Eich, W. Freude, S. Fan, Z. Yu, R. Baets, M. Popović, A. Melloni, J. D. Joannopoulos, M. Vanwolleghem, C. R. Doerr, and H. Renner, "What is — and what is not — an optical isolator," *Nature Photon.* **7**, 579–582 (2013). https://doi.org/10.1038/nphoton.2013.185
20. V. G. Kravets, A. V. Kabashin, W. L, Barnes, and A. N. Grigorenko, "Plasmonic Surface Lattice Resonances: A Review of Properties and Applications," *Chem. Rev.* **118**, 5912–5951 (2018). https://doi.org/10.1021/acs.chemrev.8b00243
21. A. K. González-Alcalde, X. Shi, R. B. Wilson, and L. T. Vuong, "Plasmonic enhancement of Faraday rotation with gold nanodisks with low height-to-diameter aspect ratios," *J. Opt. Soc. Am. B* **41**, 2480–2487 (2024). https://doi.org/10.1364/JOSAB.525700
22. B. Sepúlveda, J. B. González-Diaz, A. Garcia-Martin, L. M. Lechuga, and G. Armelles, "Plasmon-Induced Magneto-Optical Activity in Nanosized Gold Disks," *Phys. Rev. Lett.* **104**, 147401 (2010). https://doi.org/10.1103/PhysRevLett.104.147401
23. B. Auguié, X. M. Bendana, W. L., Barnes, and F. J. G. de Abajo, "Diffractive arrays of gold nanoparticles near an interface: Critical role of the substrate," *Phys. Rev. B* **82**, 155447 (2010). https://doi.org/10.1103/PhysRevB.82.155447
24. S. Murai, M. A. Verschuuren, G. Lozano, G. Pirruccio, S. R. K. Rodriguez, and J. Gómez Rivas, "Hybrid plasmonic-photonic modes in diffractive arrays of nanoparticles coupled to light-emitting optical waveguides," *Opt. Express* **21**, 4250–4262 (2013). https://doi.org/10.1364/OE.21.004250
25. B. B. Rajeeva, L. Lin, and Y. Zheng, "Design and applications of lattice plasmon resonances," *Nano Research* **11**, 4423–4440 (2018). https://doi.org/10.1007/s12274-017-1909-4
26. A. V. Baryshev, H. Uchida, and M. Inoue, "Peculiarities of plasmon-modified magneto-optical response of gold-garnet structures," J. Opt. Soc. Am. B 30, 2371-2376(2013). https://doi.org/10.1364/JOSAB.30.002371
27. M. Kataja, T. K. Hakala, A. Julku, M. J. Huttunen, S. van Dijken, and P. Torma, " Surface lattice resonances and magneto-optical response in magnetic nanoparticle arrays," *Nat. Commun.* **6**, 7072 (2015). https://doi.org/10.1038/ncomms8072
28. S. Trudel, "Unexpected magnetism in gold nanostructures: making gold even more attractive," *Gold Bull* **44**, 3–13 (2011). https://doi.org/10.1007/s13404-010-0002-5
29. A. Hernando, A. Ayuela, P. Crespo, and P. M. Echenique, "Giant diamagnetism of gold nanorods," *New J. Phys.* **16**, 073043 (2014). https://doi.org/10.1088/1367-2630/16/7/073043

30. C. K. Lim, M. J. Cho, A. Singh, Q. Li, W. J. Kim, H. S. Jee, K. L. Fillman, S. H. Carpenter, M. L. Neidig, A. Baev, M. T. Swihart, and P. N. Prasad, "Manipulating Magneto-Optic Properties of a Chiral Polymer by Doping with Stable Organic Biradicals," Nano Lett. 16 (9), 5451−5455 (2016). https://doi.org/10.1021/acs.nanolett.6b01874
31. Z. Nelson, L. Delage-Laurin, M. D. Peeks, and T. M. Swager, "Large Faraday Rotation in Optical- Quality Phthalocyanine and Porphyrin Thin Films," J. Am. Chem. Soc. 143 (18), 7096−7103(2021). https://doi.org/10.1021/jacs.1c02113
32. H. Zhu, M. Gao, C. Pang, R. Li, L. Chu, F. Ren, and W. Qin, "Strong Faraday Rotation Based on Localized Surface Plasmon Enhancement of Embedded Metallic Nanoparticles in Glass," *Small Science,* **2**, 2100094 (2022). https://doi.org/10.1002/smsc.202100094
33. OHC. Cheng, DH. Son, and M. Sheldon, "Light-induced magnetism in plasmonic gold nanoparticles," Nature Photonics 14, 365-368 (2020). https://doi.org/10.1038/s41566-020-0603-3
34. F. Royer, B. Varghese, E. Garnet, S. Nevue, Y. Jourlin, and D. Jamon, "Enhancement of Both Faraday and Kerr Effects with an All-Dielectric Grating Based on a Magneto-Optical Nanocomposite Material," *ACS Omega* **5**, 2886–2892 (2020). https://doi.org/10.1021/acsomega.9b03728
35. A. B. Khanikaev, A. V. Baryshev, A. A. Fedyanin, A. B. Granovsky, and M. Inoue, "Giant enhancement of Faraday rotation due to electromagnetically induced transparency in all-dielectric magneto-optical metasurfaces," *Opt. Express* **15**, 6612 (2017). https://doi.org/10.1364/OL.43.001838
36. D. Floess, M. Hentschel, T. Weiss, H. U. Habermeier, J. Jiao, S. G. Tikhodeev, and H. Giessen, "Plasmonic Analog of Electromagnetically Induced Absorption Leads to Giant Thin Film Faraday Rotation of 14°," Phys. Rev. X 7, 021048 (2017). https://doi.org/10.1103/PhysRevX.7.021048
37. A. Lopez-Ortega, M. Takahashi, S. Maenosono, and P. Vavassori, "Plasmon induced magneto-optical enhancement in metallic Ag/FeCo core/shell nanoparticals synthesized by colloidal chemistry," Nanoscale 10, 18672-18679(2018). https://doi.org/10.1039/C8NR03201G
38. N. Maccaferri, L. Bergamini, M. Pancaldi, M. K. Schmidt, M. Kataja, S. V. Dijken, N. Zabala, J. Aizpurua, and P. Vavassori, "Anisotropic Nanoantenna-Based Magnetoplasmonic Crystals for Highly Enhanced and Tunable Magneto-Optical Activity" Nano Lett. 16, 2533-2542 (2016). https://doi.org/10.1021/acs.nanolett.6b00084
39. H. Uchida, Y. Masuda, R. Fujikawa, A. V. Baryshev, and M. Inoue, "Large enhancement of Faraday rotation by localized surface plasmon resonance in Au nanoparticles embedded in Bi:YIG film," *J. Magn. Magn. Mater.* **321**, 843–845 (2009). https://doi.org/10.1016/j.jmmm.2008.11.064
40. H. Uchida, Y. Mizutani, Y. Nakai, A. A. Fedyanin, and M. Inoue, "Garnet composite films with Au particles fabricated by repetitive formation for enhancement of Faraday effect," *J. Phys. D* **44**, 064014 (2011). https://doi.org/10.1088/0022-3727/44/6/064014
41. A. V. Baryshev, and A. M. Merzlikin, "Tunable plasmonic thin magneto-optical wave plate," *J. Opt. Soc. Am. B* **33**, 1399–1405 (2016). https://doi.org/10.1364/JOSAB.33.001399
42. F. Zhang, T. Atsumi, X. Xu, S. Murai, and K. Tanaka, "Tunable Faraday rotation of ferromagnet thin film in whole visible region coupled with aluminum plasmonic arrays," *Nanophotonics* **11**, 275–288 (2021). https://doi.org/10.1515/nanoph-2021-0327
43. K. Sato, and T. Ishibashi, "Fundamentals of Magneto-Optical Spectroscopy," *Front. Phys.* **10**, 946515 (2022). https://doi.org/10.3389/fphy.2022.946515
44. S. Vandendriessche, and T. Verbiest, "Photoelastic modulator non-idealities in magneto-optical polarization measurements," *Polarization Science and Remote Sensing VI* 88730Z SPIE (2013). https://doi.org/10.1117/12.2022211

45. P. B. Johnson, and R. W. Christy, "Optical Constants of the Noble Metals," *Phys. Rev. B* **6**, 4370–4379 (1972). https://doi.org/10.1103/PhysRevB.6.4370

Supplementary Information
**Exceptionally high Verdet constant in gold nanodisc arrays**
*Gajendra Mulay*[1,2], *Shraddha Choudhary*[1,3], *Ashwin A Tulapurkar*[4], *Daria O Ignatyeva*[5], *Vladimir I Belotelov*[5], *Shriganesh S Prabhu*[1], and *Venu Gopal Achanta*[1,6]
[1]Department of Condensed Matter Physics and Materials Science, Tata Institute of Fundamental Research, Mumbai 400005, India.
[2]Center for Research in Nanotechnology and Science, Indian Institute of Technology Bombay, Mumbai 400076, India.
[3]Institute of Physics and Center for Nanotechnology, University of Münster, Münster 48149, Germany.
[4]Department of Electrical Engineering, Indian Institute of Technology Bombay, Mumbai, 400076, India.
[5]Faculty of Physics, Lomonosov Moscow State University, Moscow, 119992, Russia.
[6]CSIR-National Physical Laboratory, Dr K.S. Krishnan Marg, New Delhi 110012, India.
Email Address: [1,2] gajendra@tifr.res.in, [1,6] achanta@tifr.res.in.

## Methods
### Fabrication:
A layer of 80 nm gold was deposited on a quartz substrate using a direct current (DC) sputtering system. A negative e-beam resist AR-N 7520.11 was spin-coated on the gold film at 4000 rpm. E-beam lithography was employed for transferring the square nanodisc array pattern onto the resist. The dose parameter was optimized to ensure high-resolution features. Subsequently, the resist was developed for 90 seconds in AR300-46 followed by a DI-water rinse for 30 seconds. Reactive Ion Etching (RIE) was utilized to etch the exposed gold region, resulting in a nanodisc pattern. The Argon plasma etching of gold was carried out for 15 minutes with an RF power of 75 W, a flow rate of 50 sccm, and a chamber pressure of $5 \times 10^{-3}$ mbar in a SENTEC SI500 ICP-RIE system. The residual resist was removed by $O_2$ plasma etching for 10 minutes at an RF power of 80W and a flow rate of 50 sccm.

### Experimental Setups:
Dispersion measurement is measured by measuring the optical transmission spectra as a function of the incident angle $\theta_{in}$. For this, the sample was positioned on a motorized rotation stage. White light, generated by a halogen lamp, was near collimated on the sample using a combination of apertures and lenses. An Ocean Optics (Flame) spectrometer measured the optical transmission spectra. The measured transmission spectra were normalized with the transmission spectra of the quartz substrate for each incident angle.

For measuring the Faraday rotation and Ellipticity, a supercontinuum laser (NKT) with a tunable wavelength from 500 nm to 1700 nm was used as a light source. The input polarization of the incident light was fixed using a Glan-Thompson polarizer with a $10^5$:1 extinction ratio. Two quartz lenses were used to focus and collect light from the sample. Magnetic fields for the measurements were generated using an electromagnet. A photoelastic modulator (PEM) is placed in the light path after the sample and is aligned to match the orientation of the input polarizer. PEM is followed by an analyzer oriented at 45° with respect to the PEM. The signal is detected by a photodiode, which is then sent for further analysis to a lock-in amplifier and digital multimeter [44]. Supplementary Fig. S6 has the Faraday rotation measurement results of quartz substrate.

## Numerical Simulations:

The FDTD method (Lumerical FDTD solutions) and Rigorous coupled-wave analysis (RCWA) method were employed to calculate the optical scattering, transmission spectra, and Faraday rotation and electromagnetic field distributions.

To obtain the transmission spectra as a function of the incident angle $\theta_{in}$, periodic boundary conditions were applied along the x- and y-directions, while PMLs (Perfectly Matched Layers) were utilized along the z-direction. A Broadband Fixed Angle Source Technique (BFAST) employing a plane wave source was utilized, where the electric field was oriented along the x-axis. The simulations utilize a fine grid, with each element measuring 10 nm. The electric fields ($|E|$) were recorded in both the x-z and the x-y planes at respective resonance wavelengths. A perfectly matched layer (PML) boundary condition along the x, y, and z directions, along with a total-field scattering field source, was used to obtain the optical properties of a single gold nanodisc.

The diagonal and off-diagonal permittivity of gold was taken into account using the Drude-Lorentz model:

$$\varepsilon_{xx} = 1 + \frac{i\omega_p^2(\gamma - i\omega)}{\omega((\gamma - i\omega)^2 + \omega_c^2)};$$

$$\varepsilon_{xy} = \frac{i\omega_p^2 \omega_c}{\omega((\gamma - i\omega)^2 + \omega_c^2)};$$

with the parameters $\omega_p = 7.8$[eV], $\gamma = 0.1$[eV], $\omega_c = 2.06 \cdot 10^{-3} B$ [eV T$^{-1}$] reported in Ref. 21 To match the experimentally measured gold properties, Johnson and Christy [45] values were tuned to get a good agreement with the experimentally measured rotation for a smooth gold film of 0.01 deg at 850 nm wavelength $\varepsilon_{xy} = -0.07 - 0.39i$.

We observed a significant enhancement of the magneto-optical properties due to the nanopatterning, such that for the nanostructures $\varepsilon_{xy}^{nano} = 10\text{Re}(\varepsilon_{xy}) + i \cdot \text{Im}(\varepsilon_{xy})$. The value for or the wavelength of 850 nm is $\varepsilon_{xy} = 0.69 - 0.20i$.

*Figure S1 Measured optical rotation and Ellipticity for the nanostructured gold for three different periods of array, and Faraday rotation of quartz substrate.*

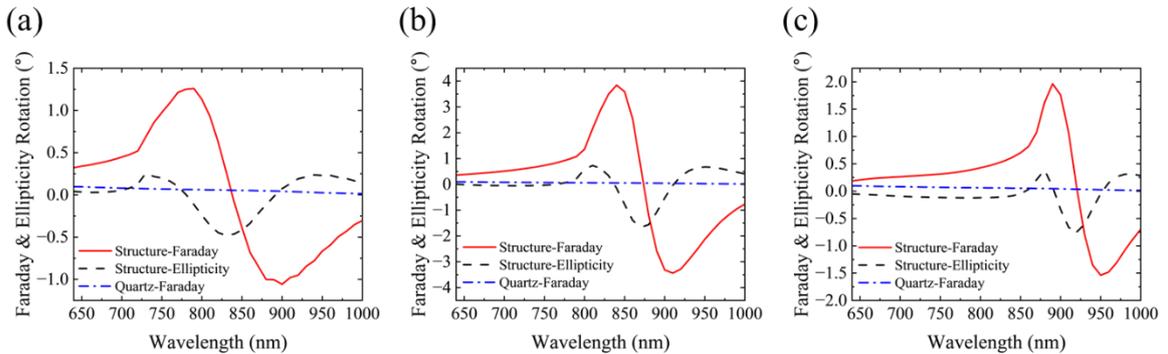

Figure S1 presents the as-measured Faraday rotation with the substrate contribution and optical rotation. The observed total rotation including both optical and magneto-optical contributions reaches a significant value of 3.8° for a 550 nm period at 900 nm. This result aligns with values commonly reported in the literature, where contributions from optical rotation and the substrate are also typically not subtracted.

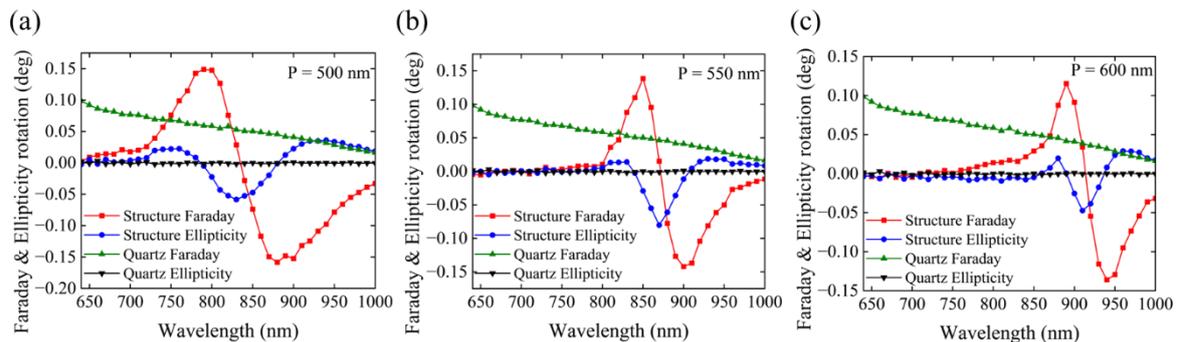

*Figure S2. Faraday rotation and ellipticity for nanostructured gold arrays with three different periods, after subtracting the contributions from the substrate and optical rotation, and Faraday rotation and Ellipticity of quartz substrate.*

Figure S2 illustrates the Faraday rotation and ellipticity values for the three different periods of the array, with contributions from the substrate and optical rotation subtracted. The maximum Faraday rotation, 0.15°, occurs at 800 nm for a period of P = 500 nm.

It should be noted that the Ellipticity ($\eta_F$) values plotted in Fig. S2 are values given by,

$$\eta_F = \eta_{TR}(B) - \eta_{TR}(0) - \eta_{TR\ SiO2}(B).$$

Where $\eta_{TR}(B), \eta_{TR}(0)$ and $\eta_{TR\ SiO2}(B)$ are the ellipticity values of structured gold (arrays) at applied magnetic field $B$, zero magnetic field ($B=0$) and the ellipticity of the quartz substrate at applied magnetic field B. So, $\eta_F$ gives the magnetic field-dependent optical rotation in the gold nanodisc arrays. Comparing the values in Fig. S1 and Fig.S2, the component of the optical rotation that is not dependent on the magnetic field is up to 1.39°. Similarly, the Faraday Rotation plotted is the field-dependent component from the gold nanodisc arrays.

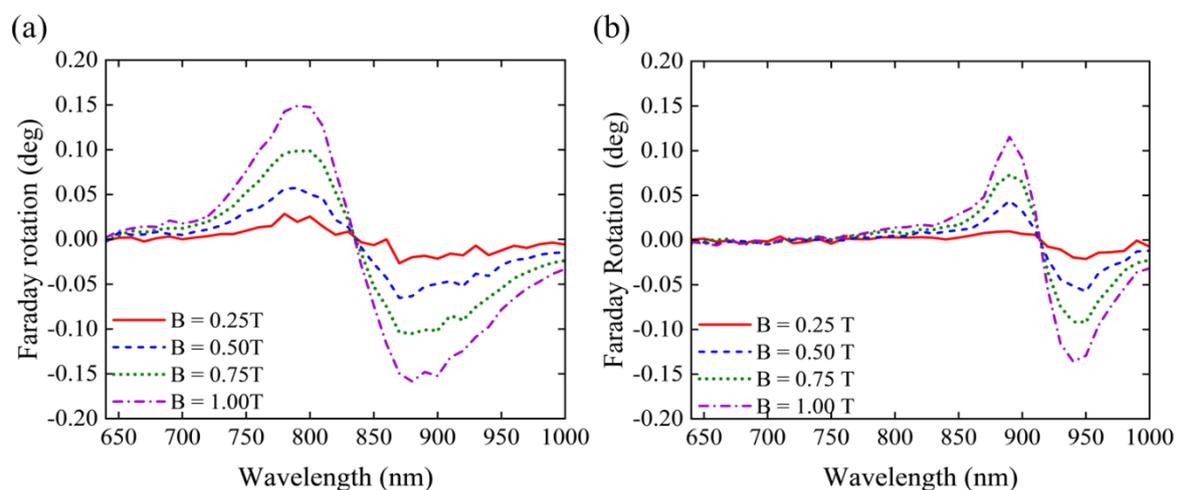

*Figure S3. Measured Faraday rotation after subtracting the substrate and field-independent contributions for different external magnetic field values for P=500 nm (a) and P= 600 nm (b) nanodisc array.*

Figure S3 depicts the measured Faraday rotation for external magnetic fields applied to arrays with periods of *P*=500 nm and *P*=600 nm. As anticipated, the Faraday rotation increases with the strength of the magnetic field and shows a resonant behaviour near the surface lattice resonance (SLR).

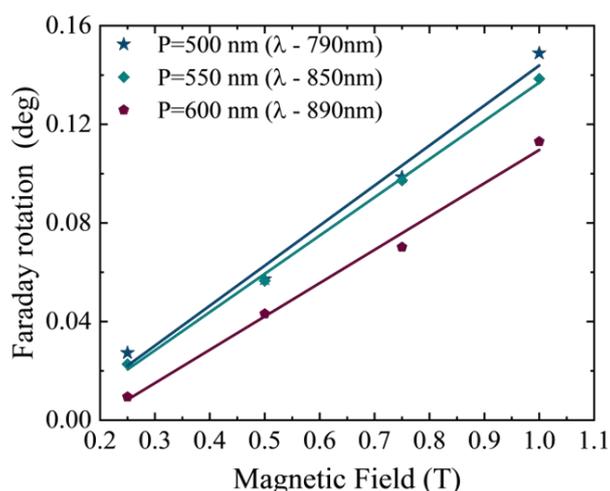

*Figure. S4 Magnetic field-dependent Faraday rotation for different wavelengths plotted from the spectra presented in Fig. S3 and in the main text.*

Purely optical rotation observed in the nanostructure might arise from fabrication imperfections resulting in the slight asymmetry of the structure. A similar effect of optical rotation in magnetoplasmonic nanostructures was previously observed in [1]. Numerical simulations show that quite small ellipticity of NPs (5-nm difference of semi-axis, in particular) results in the appearance of optical rotation with the magnitude close to the one observed in the experiment, see Fig.S5.

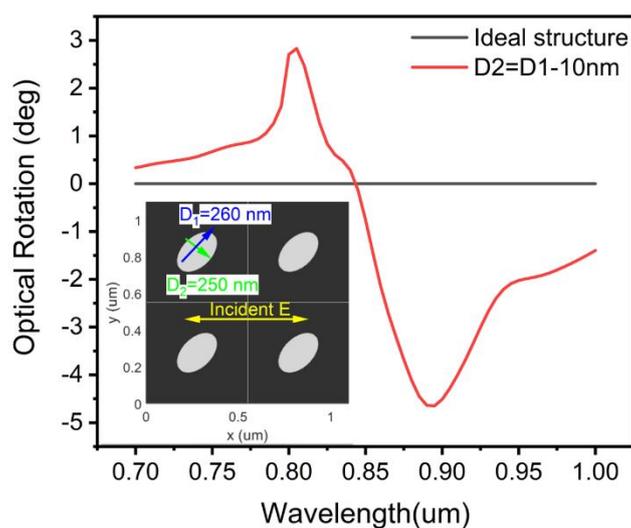

*Figure. S5 Numerically simulated optical rotation spectra observed for non-magnetized gold nanoparticles of ideal circular (Dx=Dy, black line) and elliptical (Dx=Dy-10nm, red line) shapes.*

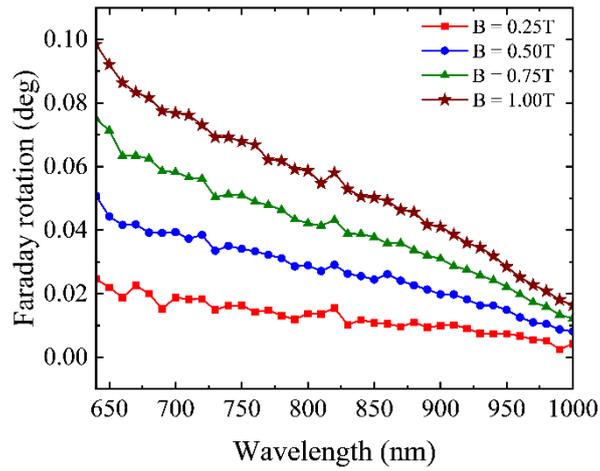

*Figure. S6 Magnetic field-dependent Faraday rotation spectra for the quartz substrate.*

## References

1. A. V. Baryshev, H. Uchida, and M. Inoue,"Peculiarities of plasmon-modified magneto-optical response of gold–garnet structures." Journal of the Optical Society of America B, 30(9), 2371-2376 (2013)